\documentclass[preprint,12pt]{elsarticle}

\usepackage{amssymb}
\usepackage{subfigure}

\usepackage{color}

\journal{Nuclear Instruments and Methods}

\begin{document}

\begin{frontmatter}



\title{The LUX Prototype Detector: Heat Exchanger Development}


\author[case]{D.~S.~Akerib}
\author[sdsmt]{X.~Bai}
\author[yale]{S.~Bedikian}
\author[llnl]{A.~Bernstein}
\author[moscow]{A.~Bolozdynya}
\author[case]{A.~Bradley}
\author[yale]{S.~Cahn}
\author[llnl]{D.~Carr}
\author[brown]{J.~J.~Chapman}
\author[case]{K.~Clark\corref{cor1}}\ead{kjc20@psu.edu}
\author[ucd]{T.~Classen}
\author[yale]{A.~Curioni}
\author[case]{C.~E.~Dahl}
\author[llnl]{S.~Dazeley}
\author[brown]{L.~de~Viveiros}
\author[case]{M.~Dragowsky}
\author[rochester]{E.~Druszkiewicz}
\author[brown]{S.~Fiorucci}
\author[brown]{R.~J.~Gaitskell}
\author[maryland]{C.~Hall}
\author[brown]{C.~Faham}
\author[ucd]{B.~Holbrook}
\author[yale]{L.~Kastens}
\author[llnl]{K.~Kazkaz}
\author[case]{J.~Kwong}
\author[ucd]{R.~Lander}
\author[maryland]{D.~Leonard}
\author[brown]{D.~Malling}
\author[tamu]{R.~Mannino}
\author[yale]{D.~N.~McKinsey}
\author[usd]{D.~Mei}
\author[ucd]{J.~Mock}
\author[harvard]{M.~Morii}
\author[yale]{J.~A.~Nikkel}
\author[case]{P.~Phelps}
\author[case]{T.~Shutt}
\author[rochester]{W.~Skulski}
\author[llnl]{P.~Sorensen}
\author[usd]{J.~Spaans}
\author[tamu]{T.~Steigler}
\author[ucd]{R.~Svoboda}
\author[ucd]{M.~Sweany}
\author[ucd]{J.~Thomson}
\author[ucd]{M.~Tripathi}
\author[ucd]{N.~Walsh}
\author[tamu]{R.~Webb}
\author[tamu]{J.~White}
\author[rochester]{F.~L.~H.~Wolfs}
\author[ucd]{M.~Woods}
\author[usd]{C.~Zhang}
\address[case]{Department of Physics, Case Western Reserve University, Cleveland, OH 44106, USA}
\address[sdsmt]{South Dakota School of Mines and technology, 501 East St Joseph St., Rapid City SD 57701, USA}
\address[yale]{Yale University, Dept. of Physics, 217 Prospect St., New Haven CT 06511, USA}
\address[llnl]{Lawrence Livermore National Laboratory, 7000 East Ave., Livermore CA 94551, USA}
\address[moscow]{National Research Nuclear University MEPHI, Faculty of the experimental and theoretical physics, Kashirskoe sh.,31, Moscow, 115409, Russia}
\address[brown]{Brown University, Dept. of Physics, 182 Hope St., Providence RI 02912, USA}
\address[ucd]{University of California Davis, Dept. of Physics, One Shields Ave., Davis CA 95616, USA}
\address[rochester]{University of Rochester, Dept. of Physics and Astronomy, Rochester NY 14627, USA}
\address[maryland]{University of Maryland, Dept. of Physics, College Park MD 20742, USA}
\address[tamu]{Texas A \& M University, Dept. of Physics, College Station TX 77843, USA}
\address[usd]{University of South Dakota, Dept. of Physics, 414E Clark St., Vermillion SD 57069, USA}
\address[harvard]{Harvard University, Dept. of Physics, 17 Oxford St., Cambridge MA 02138, USA}
\cortext[cor1]{Corresponding author}

\begin{abstract}
The LUX (Large Underground Xenon) detector is a two-phase xenon Time Projection Chamber (TPC) designed to search for WIMP-nucleon dark matter interactions.  As with all noble element detectors, continuous purification of the detector medium is essential to produce a large ($>$1ms) electron lifetime; this is necessary for efficient measurement of the electron signal which in turn is essential for achieving robust discrimination of signal from background events.  In this paper we describe the development of a novel purification system deployed in a prototype detector.  The results from the operation of this prototype indicated heat exchange with an efficiency above 94\% up to a flow rate of 42 slpm, allowing for an electron drift length greater than 1 meter to be achieved in approximately two days and sustained for the duration of the testing period.
\end{abstract}

\begin{keyword}
Noble-liquid detectors \sep Charge transport and multiplication in liquid media \sep Large detector systems for particle and astroparticle physics
\end{keyword}

\end{frontmatter}


\section{Introduction}
\label{sec:intro}

Evidence continues to mount that roughly 23\% of the energy density of the universe is constituted of a type of matter which is not baryonic in nature \cite{bertone}.  This is commonly referred to as ``dark matter'', and has become the focus of many research programs.   One of the most popular particle candidates for  dark matter is the Weakly Interacting Massive Particle, or WIMP.  Since one of the defining qualities of dark matter is that it does not interact electromagnetically, its detection requires complex instruments designed to detect rare interactions with great precision.

The LUX dark matter project \cite{lux} intends to detect WIMPS using a two phase liquid-gas xenon TPC detector \cite{doke}. Xenon, which is a scintillator, can be used to detect the small energy deposit caused by the elastic recoil of a nucleus following a collision with particles such as the hypothesized WIMPs.  The use of xenon in both liquid and gas phase allows the detection of both the scintillation and ionization electrons produced in this interaction.  This has been extensively reviewed in Ref. \cite{gaitskell}.

In addition to providing three dimensional position reconstruction for the interaction, the amount of light observed in the prompt scintillation (S1) and electroluminescence (S2) signals may be used to determine the nature of the event.  The ratio of the amount of light in the different scintillation signals changes due to the nature of the energy loss in the xenon for electromagnetically or weakly-interacting particles.

In order to maintain the level of purity required to drift these electrons in the detector medium, constant circulation of the xenon through a purification system is required.  Currently this system necessitates that the xenon be purified in a gaseous state, requiring the circulation system to evaporate and condense the xenon.  In order to mitigate the heat load imposed by these phase changes a heat exchanger was developed which allows for thermal transfer between the boiling and condensing xenon.

This paper will discuss the prototype detector constructed to develop this heat exchanger as well as to monitor the efficiency achieved.  Following this, the instrumentation used to verify the operation of the exchanger will be reviewed.  In order to show the effectiveness of purification, measurements of the electron drift length in the prototype detector will be discussed.

\section{Motivation}
\label{sec:motivation}

Arguably the biggest technical challenge for the two-phase xenon technology is achieving meter-scale electron drift lengths in the liquid (note that the LUX drift region is 50 cm deep). To date this has been based on gas phase purification through a getter, but this process is limited by the need to provide cooling to condense the returning purified xenon gas stream. In practice, this has been provided by the cryogenic systems used to cool the detector, and has resulted in fairly low flow rates - for instance, XENON100, with 160 kg xenon, has a purification flow rate of $\sim$ 5 slpm, or $\sim$ 42 kg/day \cite{xenon100}.   Larger detectors will require significantly higher gas circulation rates, and thus significantly higher cooling power.  LUX has a of goal of purifying at least 50 slpm (424 kg/day), so that the entire 375 kg xenon in the detector is passed through the purification system more than once per day.

The cooling system for both the LUX detector and the LUX prototype discussed in this paper is based on a series of liquid nitrogen thermosyphons. These thermosyphons, discussed in detail in Ref. \cite{thermosyphon_ieee}, can deliver extraordinarily high cooling power capacity over a long distance, and could in fact deliver the 450 W needed to condense a 50 slpm xenon gas stream.  However managing this much cooling power (and the equivalent heat which must be supplied for the evaporating stream) is inconvenient, will lead to large thermal gradients in the detector space, and will require a logistically challenging and expensive amount of liquid nitrogen (240 liters/day are needed for 450 W cooling). These issues are especially acute in an underground setting. Moreover, since gas purification can be readily increased to rates beyond 50 slpm, the problem of cooling is the primary challenge to achieving even more powerful purification.

We have used the LUX 0.1 prototype platform to develop a new heat exchanger technology to force a transfer of the power between the condensing and evaporating xenon streams. This development was highly successful, resulting in nearly complete heat exchange while circulating up to a gas-system limited flow rate of 42 slpm. The heat exchanger is coupled to a system which manages the path of liquid circulation through the detector and sets the height of the liquid level.

An extensive set of instrumentation was also developed to characterize both the heat exchanger and the liquid circulation system. Using the two-phase detector housed in the LUX 0.1 system, we were able to demonstrate the use of the system for purifying xenon, and in the 55 kg of xenon in LUX 0.1 achieved attenuation lengths of over a meter in an unprecedented time of only $\sim$~2 days.  The design of the heat exchanger, circulation system, and instrumentation also successfully met the dual challenges of being compatible with the limited space available in the LUX detector vessel, and being constructed of materials suitably low in radioactivity, culminating in the final system now deployed in LUX.

LUX proposed this basic technology and discussed it at conferences in 2009 and 2010 as reported in \cite{Leonard2010}. This paper includes the first full description of this work. The XENON100 collaboration has recently reported on a version of this technology which was tested at rates up to 13 slpm, but without a measurement of xenon purity \cite{columbia}.  In addition, a recent paper from the MEG collaboration details the creation of a heat exchanger with very high efficiency \cite{MEGpaper}.

\section{LUX 0.1}
\label{sec:detector}

The LUX collaboration operated a prototype detector in a surface lab at Case Western Reserve University from 2007 to 2010.  The primary goal of this preliminary version of the detector was to test the xenon circulation system designed for the final detector.  The prototype detector was referred to as LUX 0.1, to distinguish it from the full dark matter search version of the detector, which is simply called LUX.

In order to use internal plumbing and circulation systems identical to those used for the full detector, the cryostat used for LUX 0.1 was the same size as that used in LUX. A key difference between the generations is that the prototype cryostat is made of stainless steel while the full detector version is made of low radioactive background titanium. Although the cryostat used was the full size of the final version, it was desirable financially to use a reduced volume of xenon, so aluminum blocks were used for displacement.  This created an active region with a usable drift length of 5 cm while reducing the amount of xenon required to approximately 55 kg instead of the full 375 kg for LUX.

\subsection{Circulation}
\label{sec:circ}

\begin{figure}[htbp]
\centering
\includegraphics[scale = .55]{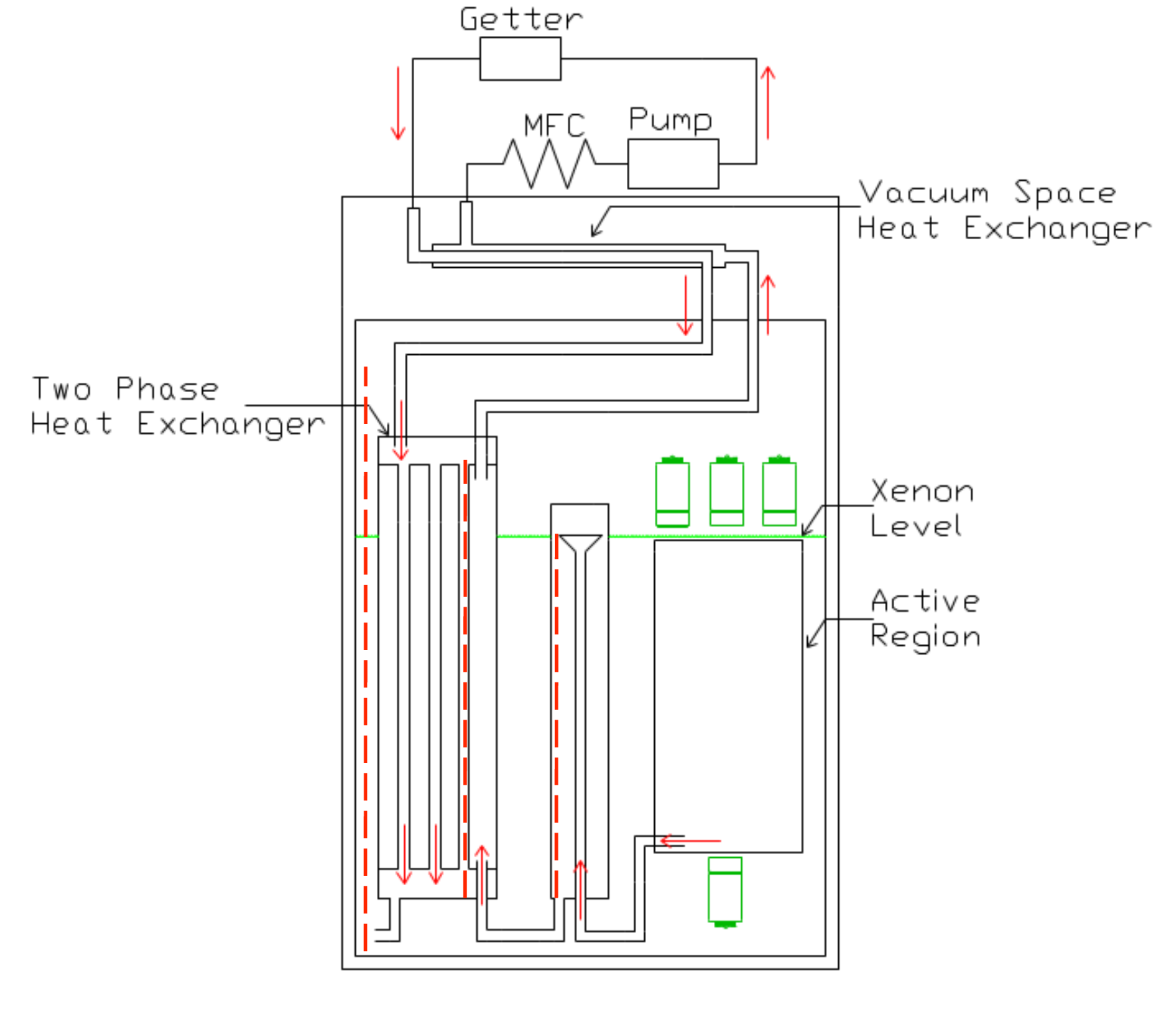}
\caption{Schematic of the circulation plumbing in LUX 0.1.  The location of the level sensors is also shown using dashed lines.}
\label{fig:lux0.1_circ}
\end{figure}

The xenon circulation path is shown in Fig. \ref{fig:lux0.1_circ}.  Gaseous xenon leaves the detector and travels through the Mass Flow Controller (MFC), which controls the flow rate using the pump to create the required pressure drop.  The xenon is then purified by passing it through a SAES PS-4 Monotorr hot metal getter and directed back into the detector.  The temperature of the incoming gas is reduced by passing it through a concentric tube heat exchanger three metres in length (located in the insulating vacuum jacket) allowing thermal contact with the outgoing gas.  This cooling of the gas reduces the heat load discussed previously in Section \ref{sec:motivation}.

The gas returning to the detector then enters the two-phase heat exchanger which divides the incoming gas into five smaller tubes.  These tubes are contained within a larger volume serving as an evaporation space for the outgoing liquid.  The goal of this heat exchanger is to efficiently transfer the large phase change energies between the incoming, condensing gas stream, and outgoing, evaporating liquid stream.  Incoming xenon, now liquified, is then emptied into the larger xenon volume, as shown in Fig. \ref{fig:lux0.1_circ}.  Xenon from this volume is drawn from the weir reservoir into the evaporator where it evaporates and is pumped out of the detector to be purified.  The design of the two-phase heat exchanger will be discussed further in Section \ref{sec:two_phase_heat_ex}.

\subsection{Instrumentation}

The instrumentation of the circulation system in LUX 0.1 consisted primarily of thermometers, heaters, and liquid level sensors.  The thermometers were platinum Resistance Temperature Detectors (RTDs), read out using a set of multiplexed Lakeshore temperature monitoring units.

Two types of heaters were employed on the LUX 0.1 chamber.  The first was part of a PID temperature control system with a maximum power output of 50 W, with the thermometer and heater both located on the vessel flange.  Two additional controllable high power electrical heaters, with maximum capacities of 400 W each, were placed on the chamber, one at the top vessel flange and one at the bottom of the chamber.  

Two varieties of capacitance liquid level sensors were used in this work: parallel wire and plate level sensors.  The parallel wire level sensors were developed for use in the circulation plumbing lines to have a minimal effect on the fluid flow through the plumbing.  Three of these level sensors were used in LUX 0.1 to measure liquid levels in the main chamber, the weir, and the condenser, and are shown as the dashed lines in Figure \ref{fig:lux0.1_schem}.

It is  advantageous to have a very accurate method of measuring the overall position of the liquid level compared to the electric field-setting grids, and, more critically, the degree to which they are parallel.  If the liquid surface and grids are not parallel, the change in field creates a non-uniform acceleration of the electrons and undesirable position dependence in the S2 signal. The measurement was accomplished with three parallel plate level sensors incorporated into the structure of the cirlex rings on the gate and anode grids (which will be discussed further in Section \ref{sec:pur_mon}) between which the capacitance was measured.  

\subsection{Two Phase Xenon Operation}
\label{sec:pur_mon}

\begin{figure*}[htbp]
\centering
\includegraphics[scale=.7]{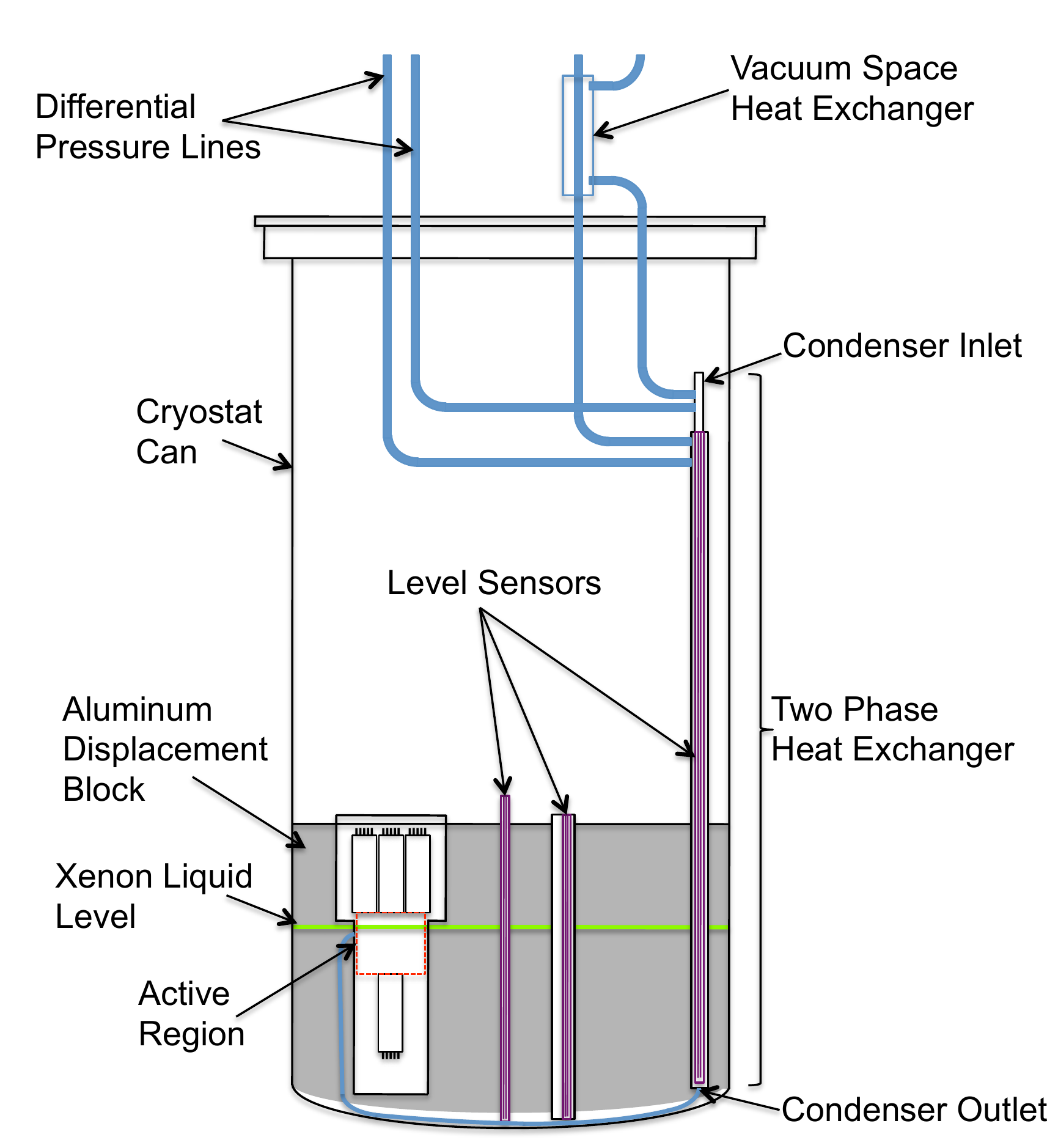}
\caption{Schematic of the LUX 0.1 prototype detector showing the aluminum displacement blocks (solid gray filled area).  Plumbing connections are also shown for reference purposes.}
\label{fig:lux0.1_schem}
\end{figure*}

The two phase xenon active region, shown in Fig. \ref{fig:lux0.1_schem}, is viewed by four 2" Hamamatsu R8778 PhotoMultiplier Tubes (PMTs), one located at the bottom (under the xenon liquid level) and three located at the top in the gas.  The plumbing connections to the heat exchanger are also shown, and will be discussed further in Section \ref{sec:two_phase_heat_ex}.

In order to drift the electrons generated by ionizing interactions in the detector volume, an electric field is applied through the active volume of the detector. In the case of the prototype detector, a series of grids created a field of up to roughly 9 kV/cm in the gaseous region and 0.44 kV/cm in the liquid volume.  The electron drift velocity at this field is roughly 1.96 mm/$\mu$s \cite{miller}.  In order to ensure these field lines remain perpendicular to the grids in the region of interest, copper rings were incorporated at evenly spaced increments along the vertical dimension.

\section{Two Phase Heat Exchanger Development}
\label{sec:two_phase_heat_ex}

\noindent A great deal of literature exists covering the mechanics of heat exchange in systems undergoing phase change.  Studying several models for local heat exchange in counter-current heat exchangers provided direction in the design of the heat exchanger used in the LUX 0.1 circulation system.  These models indicate that  in order to achieve efficient heat exchange, the surface area available to the heat exchange process per unit volume should be maximized.

Our design achieves a large surface area by using several stainless steel tubes in which the incoming xenon can condense inside a larger volume in which the outgoing xenon can evaporate.  The space constraints for LUX were also taken into account in the design of the LUX 0.1 heat exchanger, restricting the total external dimensions  to a footprint which is 3 inches wide and 0.75 inches deep.  Stainless steel was chosen due to the radiopurity achievable as well as the reduction in heat conduction along the tubes.

\begin{figure}[htbp]
\centering
\includegraphics[scale = .3]{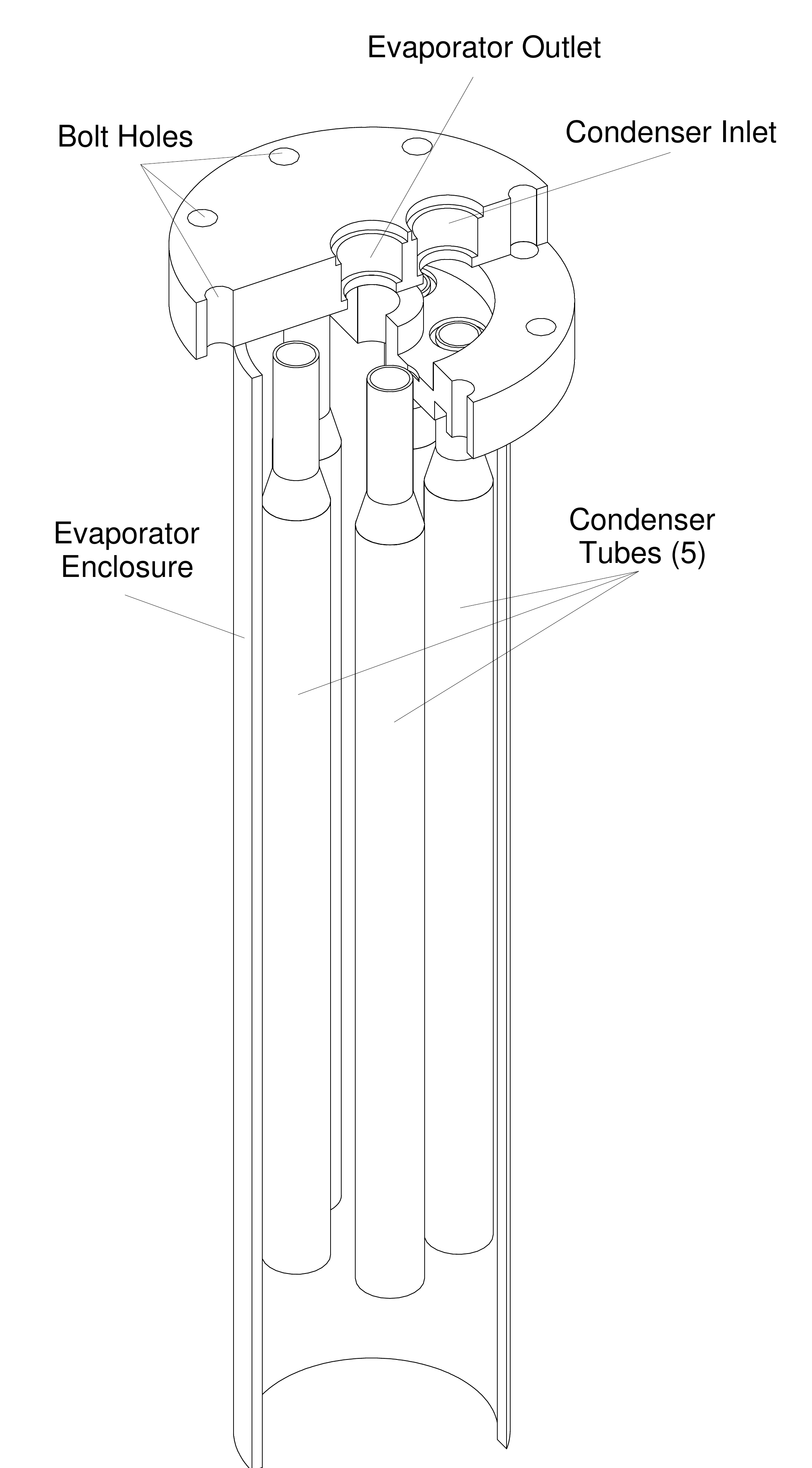}
\caption{A sketch of the top of the two phase heat exchanger as implemented in LUX 0.1.  The bottom gathers the tubes in a similar fashion.}
\label{fig:0p1_he_top}
\end{figure}

The LUX 0.1 two phase heat exchanger is shown in Fig. \ref{fig:0p1_he_top}. Xenon gas returning from purification passed through an inlet port to a set of five 1/4 inch diameter condenser lines which expanded to 3/8 inch diameter in the evaporator volume.  Liquid drawn from the detector passed through the evaporator volume surrounding the condenser tubes and exited through an outlet port.  The height of the heat exchanger was 76 cm (30 inches) which is slightly shorter than the 1 meter height available in LUX.  This length provided enough vertical room over the liquid surface to pull the effective evaporation surface up to maintain a stable evaporation/condensation interface.   Although the LUX heat exchanger is rectangular in cross section to fit within the space constraints, the LUX 0.1 heat exchanger was made of round tubing for simplicity of construction, and the cross sectional areas in the condenser and evaporator were chosen to match those available in LUX.

\begin{figure}[htbp]
\centering
\includegraphics[scale = .4]{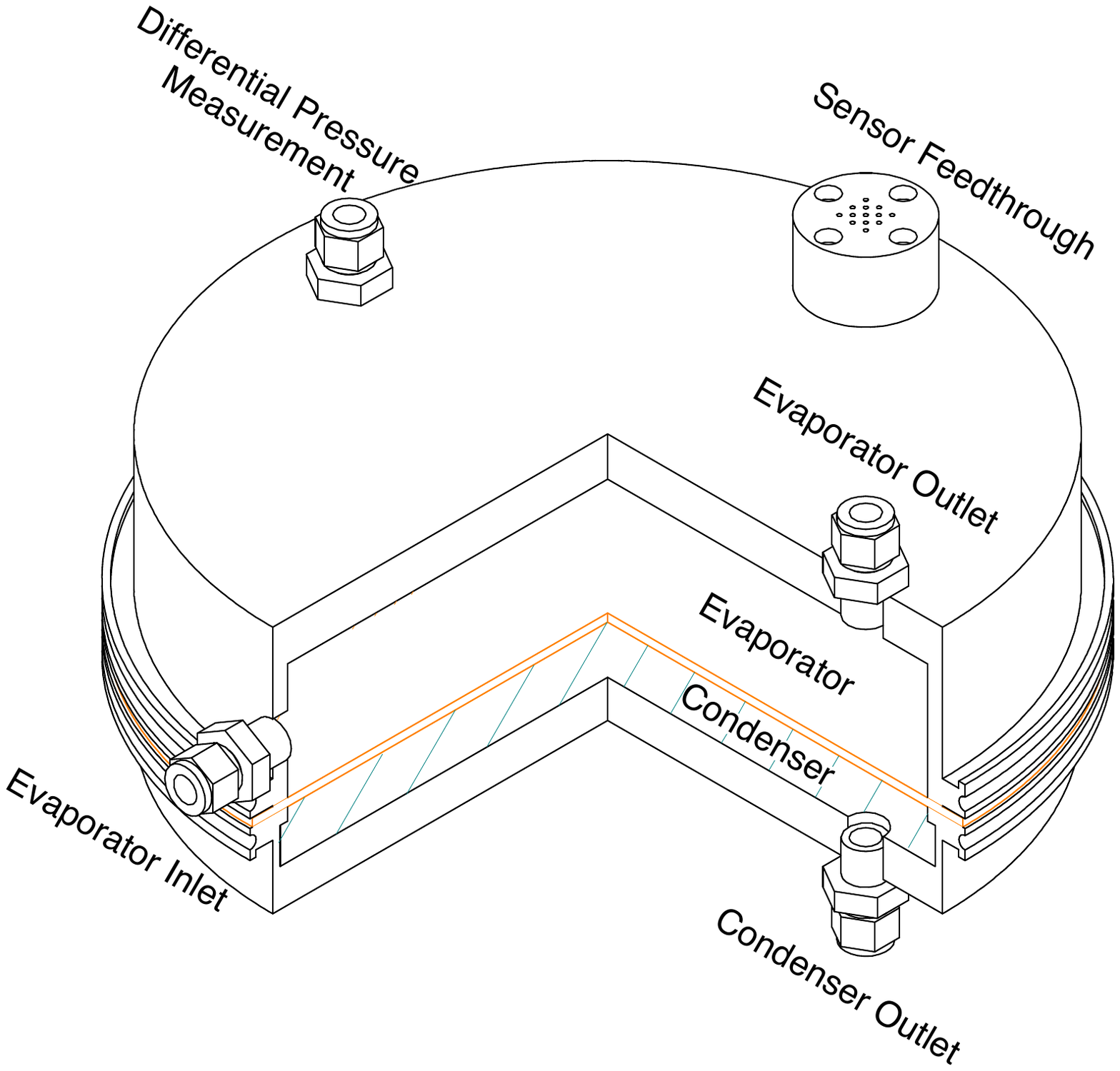}
\caption{Sketch of the ``parallel plate'' heat exchanger as deployed in early runs of LUX 0.1.  A copper membrane separates the evaporator from the condenser.}
\label{fig:par_plate_he}
\end{figure}

We also mention here an earlier version of the two phase heat exchanger based on a parallel plate design (that was run in the prototype detector with limited success) as shown in Figure \ref{fig:par_plate_he}.  The efficiency of this exchanger was measured to be higher than 90\%, up to flow rates of 20 slpm (170 kg/day).  However, at flow rates higher than this the energy transfer effectively "turned off", because the evaporation surface moved up out of the heat exchanger.  While larger flow rates could presumably be achieved with a simple scaling of the surface area in this design, the vertical tube geometry was determined to have a larger capacity in a comparatively smaller footprint.  

\section{Heat Exchanger Operation Results}
\label{sec:heat_load}

During the operation of the prototype detector, two separate methods were used to quantify the effectiveness of the two phase heat exchanger.  The first was to measure the heat load applied to the detector while circulating at several different flow rates, while the second was to identify the location of the evaporation surface during circulation.  If the two phase heat exchanger is operating effectively at the target flow rate, the evaporation surface will be contained within the heat exchanger.

\subsection{Expected Heat Load}
\label{sec:exp_heat_load}

As discussed in Section \ref{sec:circ}, purification of the xenon occurs outside of the detector in the gas phase with the purified gas being returned to the detector at approximately room temperature.  Upon leaving the purification system the xenon gas is first passed through the vacuum space heat exchanger, which cools the incoming gas from 298 K to 178 K, requiring a power transfer of 20.6 kJ/kg.  Following this first heat exchanger, the incoming gas is sent to the two-phase heat exchanger.  Condensation of the xenon at 178 K requires heat transfer of 92.5 kJ/kg, or 9.08 W/slpm (1.07 W/kg/d).  At the 50 slpm circulation target, the projected heat load due to circulation will be 454 W. The efficiency of the two-phase heat exchanger is then judged by measuring the heat load applied to the detector while circulating and comparing it with expected heat load for the measured flow rate.  There are several potential causes for inefficiencies.    One is the chance that some amount of returning xenon is not condensed in the two-phase heat exchanger.  There would then be incomplete heat exchange with the warm xenon entering the detector in the vacuum space heat exchanger.  An equivalent situation exists if there is incomplete heat exchange with evaporated xenon leaving the detector.

\subsection{Heat Load Determination}
\label{sec:heat_load_testing}

The primary method of determining the efficiency of the heat exchanger is to directly measure the additional heat load on the detector from Xe circulation.  The challenge is to identify this additional heat load in the complicated and interconnected detector system, especially because it has a number of very large thermal time constants. We begin by noting that the sum of all powers into the system, if not zero, will lead to varying temperatures of the detector components:

\begin{equation}
\sum_i C_i \dot{T}_i = \sum_k P_k
\label{eq:thermal_a}
\end{equation}

\noindent Here $k$ runs over all sources of power, while $i$ runs over all the components of the detector.  We have taken the heat capacities to be constant over the limited temperature range (roughly 5 K) of the measurements discussed in this section.  The power terms include both heat loads (for which $P_k$ is positive) and cooling provided by the thermosyphons (for which $P_k$ is negative).  While we did not measure the temperature of all of the components of the detector, there were thermometers on all the major pieces with large heat capacities, including the vessel's large top flange, the chamber wall, the Al displacer blocks, and the liquid Xe.  These components are summarized in Table \ref{tab:heat_capacity}.

\begin{table}[htb]
	\caption{Primary components of the LUX 0.1 detector and their heat capacities.  The heat capacity of the vessel flange is the sum of the two components.  Note that the remainder of the prototype detector consists of 115 kg of material.}	
\begin{center}
	\begin{tabular}{| c | c | c | c |} \hline
	Item & Material & Mass & Heat\\
	 & & (kg) & Capacity\\
	 & & & (J/K)\\ \hline \hline
	Displacement &  &  &\\
	Blocks & aluminum & 230 & 1.82$\times$10$^{5}$\\ \hline
	Vessel & stainless steel & 162 & 6.32$\times$10$^{4}$\\
	 Flange & copper & 47.2 & 1.62$\times$10$^{4}$\\ \hline
	IR Shield & copper & 97.5 & 3.35$\times$10$^{4}$\\ \hline
	Cryostat & stainless & &\\
	Can & steel & 100 & 5.10$\times$10$^{4}$\\ \hline
	Xenon & xenon & 55 & 8.69$\times$10$^{3}$\\ \hline \hline
	\bf{Total} & & \bf{691.7} & \\ \hline
	
	\end{tabular}
	\end{center}

	\label{tab:heat_capacity}
\end{table}

The power terms consist of the following. There was a ``parasitic'' heat load from the mechanical links to the room temperature vacuum vessel, blackbody radiation, and power from the residual gas pressure in the vacuum space.  Of these, only the gas pressure could potentially vary with time, but this was monitored and maintained at a level which was sufficiently low so that the parasitic power was constant throughout the measurements described in this paper. The cooling power was controlled by the amount of nitrogen in the thermosyphons \cite{thermosyphon_ieee}, which was held constant throughout these measurements.  The parasitic and cooling powers, while large, are of opposite sign and thus largely cancel.  There are two other powers, which varied during our measurements.  The first was the electrical power from the PID temperature controller, used to control the temperature of the detector, the value of which was recorded directly from the temperature controller.  The second was the heat load from circulation, which the heat exchanger seeks to minimize. 

Listing these terms explicitly, the circulation power can now be written as follows from Equation \ref{eq:thermal_a}:

\begin{equation}
P_{circulation} = \sum_i C_i \dot{T}_i  - P_{PID~heater} - P_{parasitic} - P_{cooling}
\label{eq:thermal_b}
\end{equation}

\begin{figure}[h!tb]
\centering
\subfigure[Effect alteration in xenon flow rate has on monitored temperatures]{
\hspace{-2.5 em}
\includegraphics[scale = 0.73, angle=0]{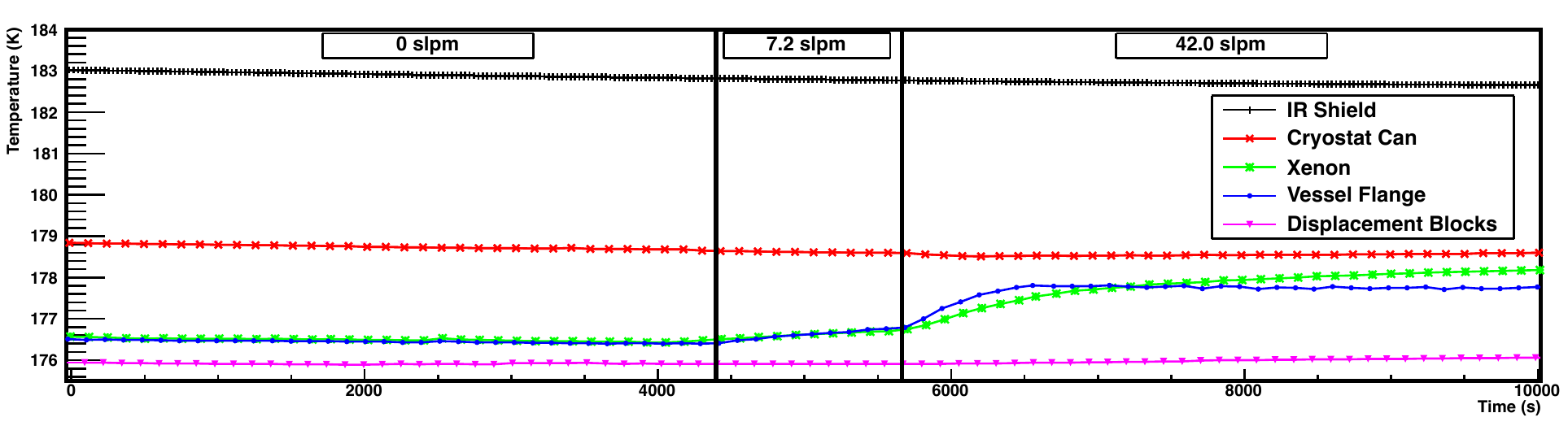}
}
\subfigure[Effect alteration in xenon flow rate has on derived powers.]{
\hspace{-2.1 em}
\includegraphics[scale = 0.73, angle=0]{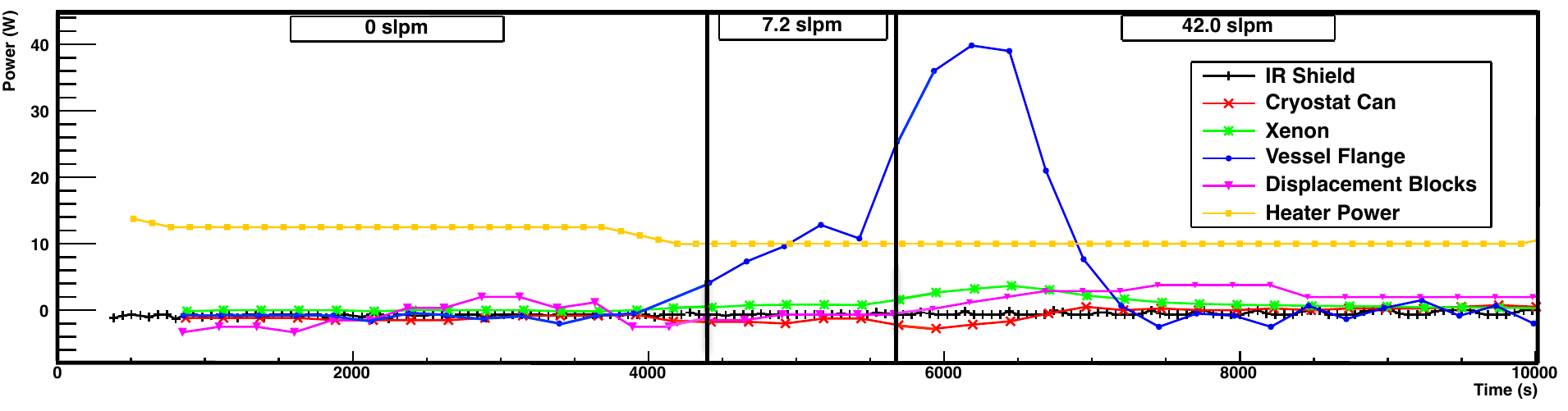}
}
\caption{Data taken during prototype operation showing in the top plot the effect of changing the xenon flow rate on the temperatures.  The bottom plot shows the smoothed instantaneous power measured using the components of the experimental setup.}
\label{fig:power_monitoring}
\end{figure}

\noindent The terms on the right hand side of Equation \ref{eq:thermal_b} have to be deduced from the data, an example of which is shown in Fig. \ref{fig:power_monitoring}.  During the time these data were taken, the xenon circulation rate was changed from 0 slpm to 7.2 slpm for a short period before it was increased to 42.0 slpm.  The top plot in Figure \ref{fig:power_monitoring} shows the temperatures of the major detector elements.  The bottom plot shows the derived power terms (C$_i \dot{T}_i$) from these temperatures and the heat capacities in Table \ref{tab:heat_capacity}, as well as the power applied by the PID temperature controller.  


The basic method of finding P$_{circulation}$ is to compare PID power and derived powers before and after circulation begins.  The parasitic and cooling powers, which are not measured directly, do not vary over this time period, and consequently cancel in this difference.   By measuring the PID power over a long period of time with no circulation and all temperature derivatives near zero, we measured the sum of the parasitic and cooling powers to be $<$ 1.5 W.  Thus, the magnitude of any slow variation in this sum power is negligible.

When the flow rate is changed, the temperatures of several components of the detector adjust.  Ideally, the data would be taken over a sufficiently long period such that all temperature derivatives, and hence derived powers, are zero, and the change in circulation power would be measured as the decrease in the PID heater power.  It is clear that this was achieved at gross level for the data in Figure \ref{fig:power_monitoring}, and that the heat load from circulation is small.  

In practice, however, not all temperatures stabilized.  In Fig. \ref{fig:power_monitoring} the displacement blocks are still warming at end of the period shown, and the $\sim$ 2.9 W positive derived power counts towards the deduced circulation power.  It is also clear in this data set that independent of changes in the xenon circulation rate, there were long-term changes in the PID heater power, and to a lesser extent, some of the temperature-derived powers.  These effects are small but not readily removed from the data.   For all data reported here, the PID power had a slight downward slope prior to increasing the flow rate. To obtain a conservative estimate of the heat exchanger efficiency we did not attempt to correct for these pre-existing trends, instead attributing these erroneous extra powers to circulation. For the example flow point of 42 slpm derived from Fig. \ref{fig:power_monitoring} the deduced circulation power is increased by a 2.6 W drop in PID power.  For most of the flow rates we did not obtain as long or as stable a period of data as shown in Fig. \ref{fig:power_monitoring}, and so the apparent power-increasing effects described here were larger, and thus the deduced heat exchanger efficiencies were lower.  The method shown here assumes the conservative approach of attributing the entire heat load to the circulation.

As a separate cross check on this basic method we applied heat loads of first 20 W and then 30  W to the vessel using the electrical heater located at the bottom of the chamber, and looked for the differential response in the detector. These large heat loads caused all elements of the detector to warm. The temperature of the aluminum displacer blocks increased steadily throughout the tests, but time was allowed for all other elements of the system to stabilize. With 20 W applied, the rate of temperature increase of the blocks was 2.630$\pm$0.002$\times$10$^{-4}$ K/s while an applied heat load of 30 W resulted in a rate of 3.17$\pm$0.03$\times$10$^{-4}$ K/s.  The difference in heating rate is then 5.43$\pm$0.03$\times$10$^{-5}$ K/s, from which the derived power is 9.83$\pm$0.19 W, in good agreement with the 10 W difference in applied powers.

\subsection{Efficiency Calculations}
\label{sec:eff_calc}

\begin{figure}[htbp]
\centering
\includegraphics[scale = .55]{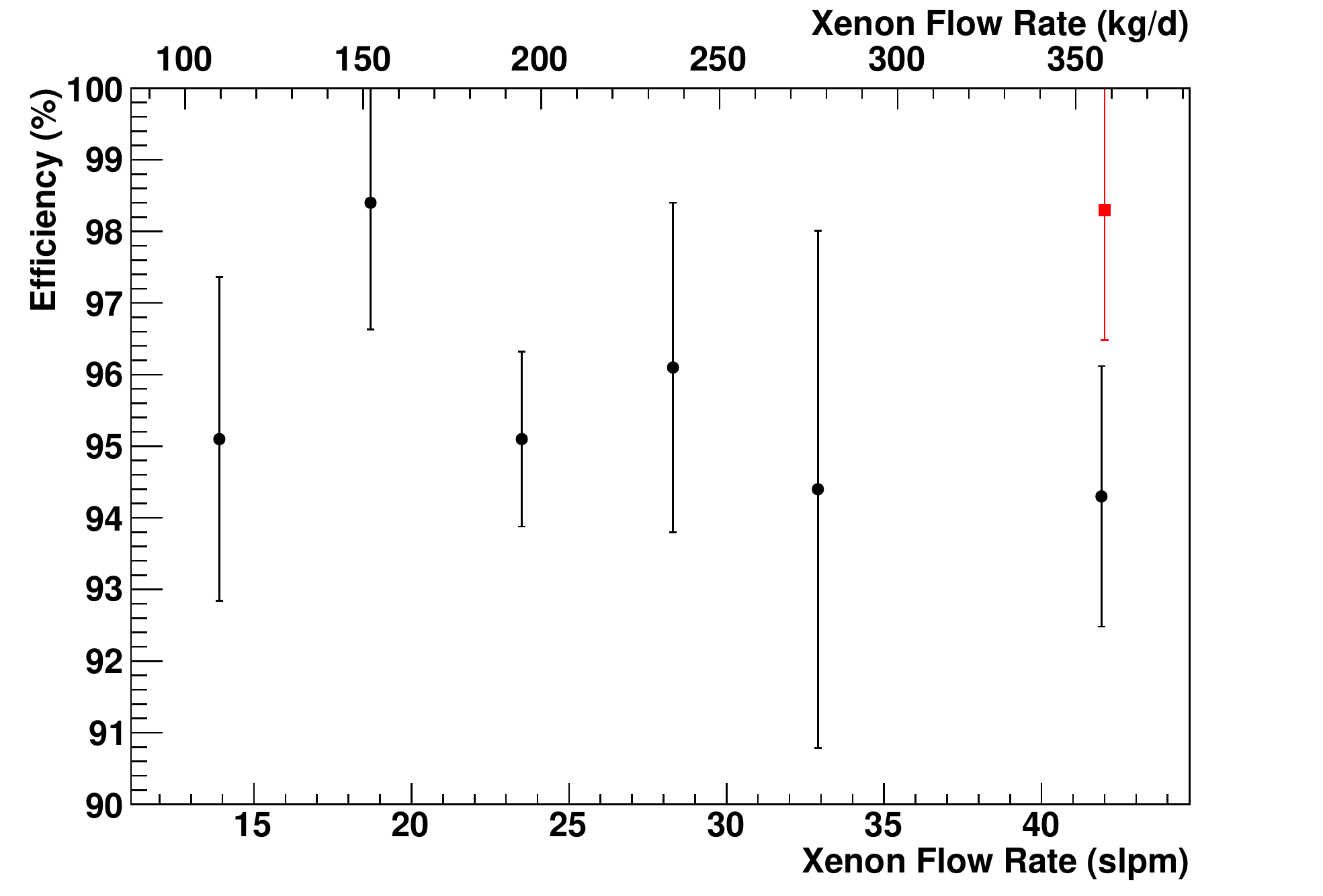}
\caption{Efficiency of heat exchanger with flow rate in the LUX 0.1 prototype detector.  Uncertainty in the flow rate is smaller than the point size and has not been shown.  The higher of the two efficiency points at 42 slpm (the red square) comes from the data shown in Fig. \protect\ref{fig:power_monitoring}. }	
\label{fig:eff_plot}
\end{figure}

Using the calculated heat loads and the measurement techniques described in Section \ref{sec:heat_load_testing}, the efficiency of the two-phase heat exchanger can be determined by dividing the measured heat load by the expected heat load, 9.08 W/slpm, plotted in Figure \ref{fig:eff_plot}.  

The error bars shown in Fig. \ref{fig:eff_plot} include the uncertainty from the fits to the temperature derivatives, and an estimated 5\% error in the flow as measured by the mass flow controller (which results in an error in the expected heat load).  The flow uncertainty directly gives an error on the x axis, though no error bar is shown for this.  An additional effect worth noting, but not addressed in this analysis, is that the heater power as reported by the PID temperature controller was restricted to 2.5 W increments.

From these measurements, we find that the efficiency of the prototype heat exchanger remains above 94\% up to the maximum flow achievable in LUX 0.1, 42 slpm (382 kg/d), as limited by the gas system.  As previously stated, the target flow for LUX is 50 slpm, which is approximately 424 kg/d.  This translates into turning over the entire mass of the detector more than once per day in LUX, compared to 6.9 times per day for the 55 kg prototype used here.  Since the design goals were achieved, this heat exchanger design was deemed a success. 

\section{Location of Evaporation and Condensation Surface}

\label{sec:evap_height}

One additional way to evaluate the heat exchanger effectiveness is to monitor the height of the liquid surface in the evaporator.  It is clear that if evaporation of the outgoing liquid happens outside of the heat exchanger, useful heat transfer, and therefore optimum efficiency, cannot be achieved.  Since the circulation lowers the pressure in the evaporator proportional to the flow rate, increasing the flow rate is expected to slowly raise the evaporation interface inside the heat exchanger.  Unfortunately, this process does not create a smooth liquid surface \cite{carey}.  This means that the usual, two-wire capacitance method of determining the liquid level was not useful in this case.

Two complementary techniques were used to determine the height of the liquid level in the evaporator in LUX 0.1.  The first of these was to measure the pressure difference between the evaporator volume and the xenon space. This hydrostatic pressure can then be used to calculate an effective liquid surface height. The second method involves instrumenting the evaporator with several thermometers distributed internally along the vertical axis.  Thermometers which are under the liquid surface report similar temperatures as convection maintains a constant temperature in the liquid.  A thermometer near the height of the boiling surface would report a correspondingly colder value than the others, as it experiences evaporative cooling.  Thermometers above the liquid surface show an increasing temperature with their height as the gaseous xenon increases in temperature with the conclusion of mist evaporation.

\begin{figure}[htbp]
\centering
\includegraphics[scale = .55]{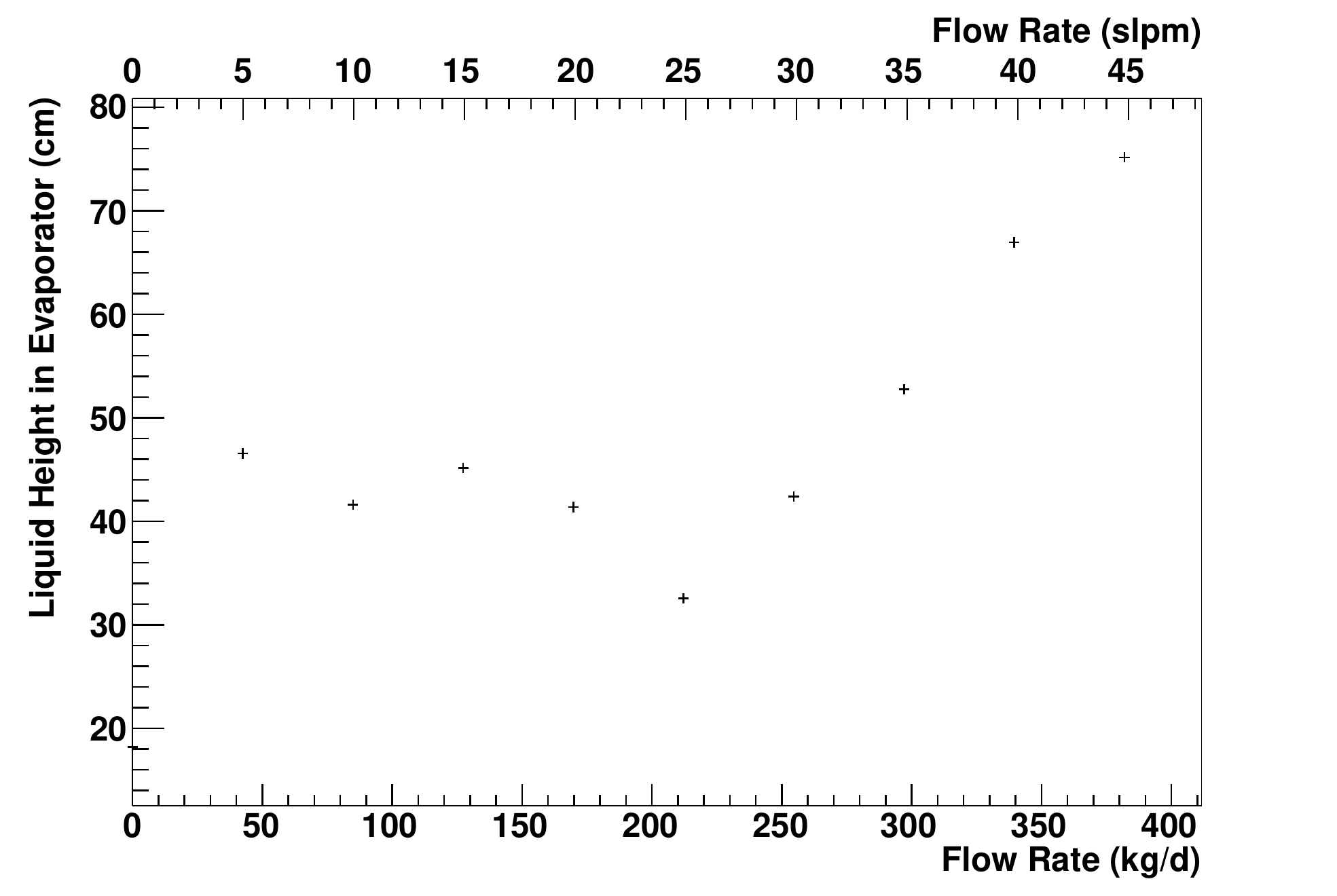}
\caption{The effective liquid height in the evaporator plotted as a function of the flow rate using the differential pressure measurement.  The height is taken relative to the height of liquid in the xenon space.  Error bars are not included in this illustrative plot.}
\label{fig:evap_height_flow}
\end{figure}

The use of both the thermometers and the differential pressure sensor provided consistent representations of the height of evaporation for the majority of the flow rates.  Since the thermometer-dependent measurement gives only discrete values due to the placement of the thermometers, the differential pressure measurement was used to determine the height.  The differential pressure results for the effective liquid height in the evaporator as the circulation rate was increased are shown in Fig. \ref{fig:evap_height_flow}.

The evaporator extended 60 cm above the xenon liquid level in the detector.  This means that when using the differential pressure measurement, the effective liquid surface appeared to extend above the evaporator at circulation rates higher than 40 slpm (340 kg/d), which should dramatically reduce the heat exchange.  This is in contradiction with the efficiency measurements, shown in Fig. \ref{fig:eff_plot}, which indicate that at circulation rates which seem to have caused the effective liquid level to leave the evaporator, efficiencies of greater than 94\% are still achieved.

The condenser was also outfitted with a level sensor to monitor the liquid level. It was  expected that this level would be the same as that in the main chamber during periods of no circulation, and would lower as the circulation rate was increased. However it was observed that for flows above approximately 5 slpm, the level sensors showed the liquid level to be at the bottom of the condenser.  This does not necessarily mean that xenon is not condensing in the condenser: we would not expect the level sensor to detect liquid condensing on the walls of the condenser tubes and trickling down.  However this does mean that we cannot be certain that there is no un-condensed xenon gas escaping from the condenser.  Such gas would then enter the xenon detector volume and condense on any cooled surface. The only surfaces with cooling power to condense xenon are the top copper flange (being cooled by the thermosyphon), the exterior of the evaporation chamber, and, if  evaporation has partially moved  out of the heat exchanger,  the outlet tube from the evaporator.   Condensation on the flange would have been detected as a heat load in the temperature and power measurements as discussed in Section 5.2 - this effect must be $<$ 5\%.   The outside of the main evaporator chamber was insulated with a PTFE tube, so there should have been no condensation there.   Figure 7 shows that for all but the highest flow rates, evaporation is confined to the heat exchanger.  This means that gas could not be condensing on the outside of the evaporator outlet tube, except for possibly some small amount at the highest flows.

\section{Xenon Purification}

\subsection{Overview}
The success of the prototype heat exchanger design allowed for continued studies of xenon purification with the LUX gas system fully implemented.  A series of runs were undertaken with the prototype detector to study xenon purification, including the one discussed in this paper, carried out at a circulation rate of 6~kg/hr (20 slpm) through a SAES Monotorr Zr heated getter. For a detailed analysis of the purification mechanism and efficiency of this getter technology in a xenon environment, see Ref. \cite{Leonard2010}.

In order to measure the impurity concentration in the xenon inside LUX 0.1, electron-drift calibrations with monoenergetic gammas from $^{57}$Co (122~keV) and $^{133}$Ba (356~keV) were performed. Photoabsorption interactions in the liquid xenon active region generate both a 178~{nm} prompt scintillation signal and ionization electrons. These scintillation signals (S1) were collected by the PMTs and used to measure the amount of energy deposited in the xenon. The ionization electrons from the photoabsorption interaction were drifted upwards with a 0.44~{kV/cm} electric field into the gaseous xenon region, where they emitted photons due to electroluminescence and produced secondary scintillation signals (S2), which were also detected by the PMTs. Since electron drift occurs at a known constant velocity for a given field, the time between the S1 and S2 signals is used to measure the vertical location of the photoabsorption, or the event depth. 

\begin{figure}
\begin{centering}
\includegraphics[width=1\columnwidth]{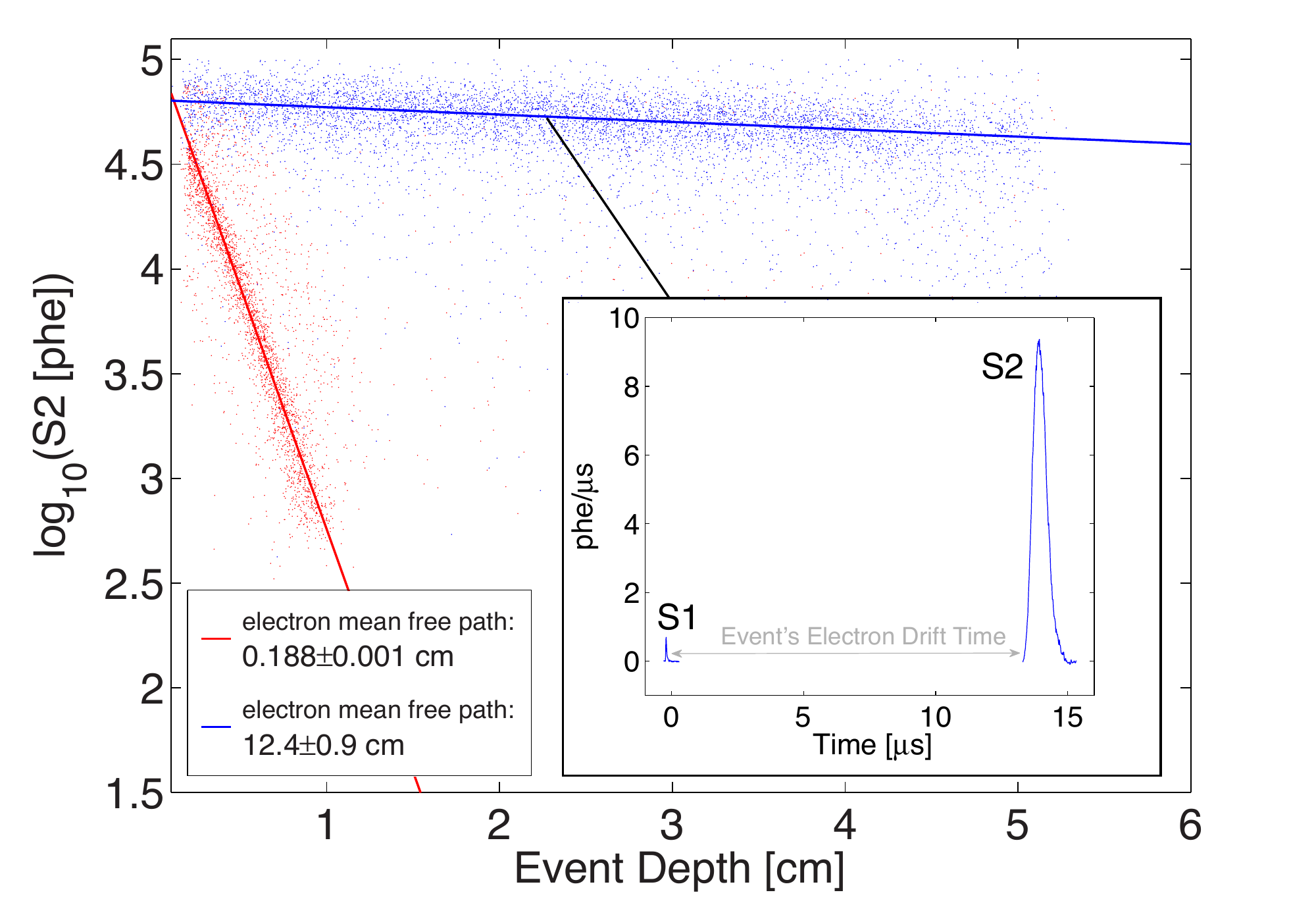}
\par\end{centering}
\caption{Comparison of electron drift lengths
for two different xenon purity levels. Red (the lower line) corresponds to a low-purity
dataset, with a measured e-folding drift length of $0.188 \pm 0.001$~cm.
Blue (the higher line) shows a relatively high-purity dataset with a measured e-folding
drift length $12.4\pm0.9$~cm. The inset shows an example event used
to monitor the electron drift length. The electron drift time between the S1 and the S2 signals can be converted to distance since the electrons move with a known velocity in liquid xenon. }
\label{driftsample}
\end{figure}

Electronegative impurities present in the liquid xenon capture electrons from the S2 signal, and this attenuation increases exponentially with event depth. One way to gauge charge loss caused by impurities in the xenon is to measure the size of the S2 signal as a function of depth to obtain the electron's effective mean free path $\lambda_e$, defined as the distance at which the S2 signal drops by a factor of $e^{-1}$. Figure \ref{driftsample} shows an example of the analysis used to measure the electron mean free path in LUX 0.1. The S2 size in photoelectrons (phe) detected with the photomultiplier tubes is plotted against event depth for all photoabsorption events from the  $^{57}$Co~122~keV line. The conversion from drift time between S1 and S2 signals to event depth is cross-calibrated by measuring the largest available drift time, which corresponds to the full length of the drift volume (5~cm). The electron effective mean free path was then obtained from the reciprocal of the slope in the linear fit of log(S2).

Low-purity xenon will display an S2 distribution similar to the lower curve in Figure \ref{driftsample}. As the xenon is purified and the number of impurities decreases, the S2 distribution flattens, as depicted by the upper curve. Experimental errors in $\lambda_e$ also increase as the S2 distribution flattens, since $\lambda_e$ is a measurement of the inverse of the slope; the errors diverge as $\lambda_e$ becomes much greater than the size of the drift region.

\subsection{Results}

\begin{figure}[hhht]
\centering
\includegraphics[width=0.65\textwidth]{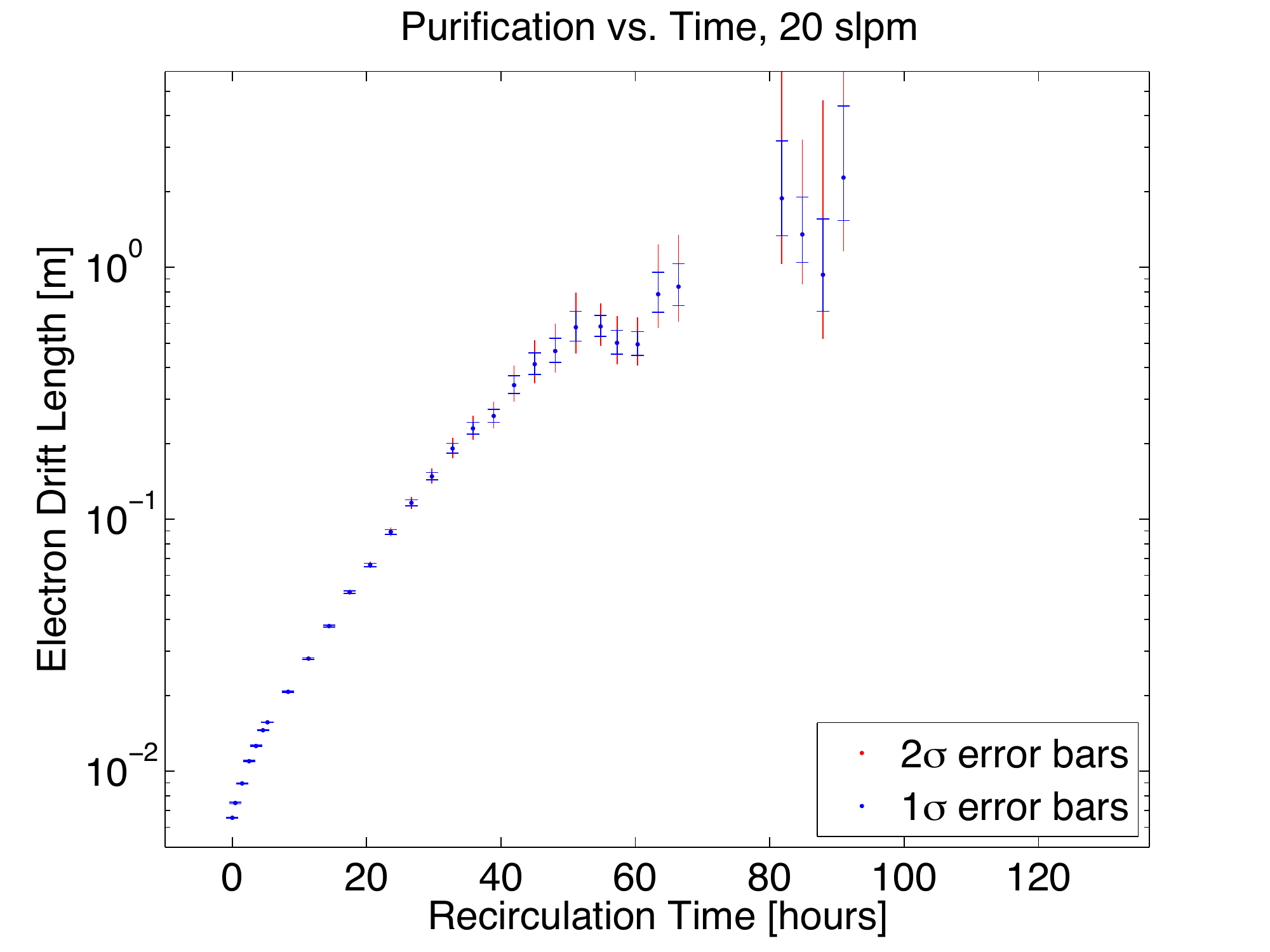}
\caption{The electron drift length has the expected behavior as a function of time, with an exponential increase at initial times with a time constant and saturation at later times dependent on circulation rate. An electron mean free path greater than 1~m was achieved at the 95\% CL. The last four datasets have large uncertainties since the measured drift lengths are much larger than the detector's drift region of 5~cm.}
\label{fig168}
\end{figure}

The electron drift length was measured over a period of more than 90 hours with the detector temperature at 167.8~K. The xenon started out dirty, with $\lambda_e = 6.61$ mm, and was continuously purified and circulated at a rate of 6.0~kg/hr. The resulting measurements of the attenuation length over time are shown in Fig. \ref{fig168}.  The drift velocity of the electrons was also calculated and shown to have a value of 1.63 mm/$\mu$s, with more details contained in Reference \cite{carlos_paper}.

An electron mean free path greater than 1~m was obtained at the 95\% CL, a measurement that was limited by the large statistical error from measuring an electron drift length that was much larger than the size of the drift region (5 cm).  The temperature dependence of the purification results were also studied with results presented in Ref. \cite{carlos_paper}.

\section{Summary}
\label{sec:summary}

The LUX 0.1 prototype successfully demonstrated the  construction and integration of the circulation and purification systems for use in LUX.  Using these systems, purification of the xenon was accomplished with a maximum drift length of greater than one meter following 50 hours of circulation at 6.0 kg/hr.  In addition, heat exchange of greater than 94\% efficiency has been shown to be achieved up to 42 slpm (382 kg/d).  These results illustrate the potential for the full LUX detector, which is expected to begin underground operation in 2012.

\section*{Acknowledgements}

This work was partially supported by the U.S. Department of Energy (DOE) under award numbers DE-FG02-08ER41549, DE-FG02-91ER40688, DOE, DE-FG02-95ER40917, DE-FG02-91ER40674, DE-FG02-11ER41738, DE-SC0006605, DE-AC52-07NA27344, the U.S. National Science Foundation under award numbers PHYS-0750671, PHY-0707051, PHY-0801536, PHY-1004661, PHY-1102470, PHY-1003660, the Research Corporation grant RA0350, the Center for Ultra-low Background Experiments at DUSEL (CUBED), and the South Dakota School of Mines and Technology (SDSMT). We gratefully acknowledge the logistical and technical support and the access to laboratory infrastructure provided to us by the Sanford Underground Research Facility (SURF) and its personnel at Lead, South Dakota.

\end{document}